\documentclass[preprint,12pt]{elsarticle}
\usepackage{xcolor}

\usepackage{amssymb}

\journal{Journal of Theoretical Physics}

\begin{document}

\begin{frontmatter}



		\title{The start of the Abiogenesis: Preservation of homochirality in proteins as a necessary
		and sufficient condition for the establishment of the metabolism }

\author{S\o ren Toxvaerd }

\address{ Department
 of Science and Environment, Roskilde University, Postbox 260, DK-4000 Roskilde, Denmark}

\begin{abstract}

Biosystems
contain an almost infinite amount of  vital important
details, which  together ensure
their life. There are, however,
some common structures and reactions in the systems:
the homochirality of carbohydrates and proteins, the metabolism and the genetics.
The Abiogenesis, or the origin of
life, is probably not a result of  a series of single events, but rather the result
of a gradual process with increasing complexity of molecules and chemical
reactions,
and the prebiotic synthesis of 
molecules  might not have left a trace  of the establishment
of   structures and reactions at the beginning of the evolution. But
alternatively, one might be able to determine some order in the 
formation of the  chemical denominators in the Abiogenesis.
Here we review experimental results and present a  model of the start of the Abionenesis, 
where the spontaneous
formation of homochirality in proteins  is the precondition  for
the establishment of homochirality of carbohydrates and for the metabolism 
at the start of  the  Abiogenesis.

\end{abstract}


\end{frontmatter}

\section{The preservation of 
homochiral structures in millions of years in a prebiotic aqueous environment.}
There have been many proposals to  the
establishment of homochirality of
carbohydrates and amino acids in a prebiotic environment. It is, however,
not the establishment, but  rather the preservation of homochirality,
which is the problem. Both the units of amino acids in the proteins, as well as the central molecules, Glyceraldehyde and  Glyceraldehyde-3-phosphate,
in the polymerization of carbohydrates
have  active isomerization kinetics \cite{Bada,Nagor}, and this  will, without other chemical constraints, give  racemic 
compositions  in  aqueous solutions  \cite{Tox3}. Biosystems are from a physicochemical point of view, soft condensed matter in 
aqueous environments, and the
Abiogenesis must have taken  millions of years, so the establishment of homochirality should
in fact be readdressed to:
$\textit{the establishment and preservation of homochiral struc-}$\\ 
$\textit{tures in millions of years in an aqueous environment.}$

The chiral composition and stability  of a system of chiral
units is determined by 
the ratio between a decrease in entropy, and thereby an increase
of Gibbs free energy by homochiral ordering,  ($\approx$ 2 kJ/mol), and the gain
in enthalpy by a chiral discrimination in favour of homochirality \cite{Tox1,Tox2,Tox3}. 
The chiral discrimination is given by the difference in enthalpy, $\Delta H_{\textrm{CD}}$,
between the enthalpy, $\Delta H_{\textrm{f}}$ of a homochiral system, $(L)$, and a corresponding
racemic mixture, $(DL)$. The chiral discrimination  must be negative and
and less than  $\div$2 kJ/mol in order overcome the loss of entropy and to establish homochiral domains.
The formation enthalpy, for  for  crystals, (c), $\Delta H_{\textrm{f}}^{\ominus}(c)$, can be determine from
combustion energy, and the chiral discrimination is
\begin{equation}
 \Delta H_{\textrm{CD}}^{\ominus}(c)= \Delta H_{\textrm{f}}^{\ominus}(c,L)-\Delta H_{\textrm{f}}^{\ominus}(c,DL).
\end{equation}

A chiral system in an aqueous environment and with isomerization kinetics and chiral discrimination of its chiral units
tends to a homochiral ordering at high concentrations of its chiral units and low water activity, but racemizes at higher water activity.
The  solution in a cell (cytosol) can be characterized as an aqueous-like solution with an emulsion of proteins, and
in a recent article  \cite{Tox3} it was argued, that the 
establishment of homochirality was started and preserved in the hydrophobic core of enzyme-like proteins.
The conclusion was obtained from
thermodynamic considerations and from Molecular Dynamics simulations of a   model of   peptide chains in aqueous 
solutions. Here we shall review and summarize experimental results which support this hypothesis.
The spontaneous formation of homochiral proteins was, however,
also   a sufficient, but necessary  condition  for
  the establishment of homochirality of the carbohydrates, and the metabolism  at the start of the  Abiogenesis.

\section{The establishment and preservation of 
homochiral structures  in proteins.}

The  structures of proteins were predicted by Linus Pauling in a series of papers in 1951, where he  derived the two
 secondary structures of proteins, the $\alpha-$helix \cite{Pauling1}
and the pleated $\beta-$sheet conformation \cite{Pauling2}. 
The  clockwise (positive) $\alpha-$helix structure was derived
from spectroscopic data for bond lengths and angles at the
planar peptide units, and with an $\it{intramolecular}$ stabilizing hydrogen-bond 
between a hydrogen atom at the substituted -N-H and an oxygen atom in a
-C=O group in the helix. The energy of the -N-H$\cdot\cdot\cdot$O=C- hydrogen bond ($\approx$ 10 kJ/mol) is only
of the order  a half of the energy of a  corresponding $\it{intermolecular}$
hydrogen bond \cite{Wendler}, and this fact has  important consequences. The loss of
intermolecular hydrogen bond energy by the establishment
of the peptide bonds results in an emulsion-like  aqueous phase separation with compact globular proteins,
where the weaker  intramolecular hydrogen bonds stabilize
the secondary  protein  structures.

The proposed  $\alpha-$helix (Figure 2 in \cite{Pauling1} ) is, however, not the  
 clockwise all L-$\alpha-$helix structure, but its mirror structure, an anticlockwise all D-helix. \cite{Dunitz} .
 This mix of notation was
 irrelevant for  Linus Pauling and for the contents of his article, but might be relevant for understanding the ability
 to maintain homochirality in the hydrophobic compact part of the globular protein.
 The two helices have the same minimum intramolecular
 energy (enthalpy), and although one cannot conclude, that helix(+)- structures with a more  racemic composition and
 with even lower energy does not exist, it is not likely.
 The enthalpy of a protein  is given by the intramolecular  energies, including the energy from the
intramolecular hydrogen bonds and the "hydrophobic forces" (surface tension) from the
loss of intermolecular hydrogen bonds by the phase separation, 
which all together ensure the compact core and a sufficient chiral discrimination for the homochirality.
However, some bioproteins with a few units of D-amino acids exist \cite{Fujii},
which indicates  that homochirality in proteins, although it stabilizes the  compact helical state \cite{Nanda}, is not a
strict necessity for the establishment of the secondary structures in proteins.
 
The active isomerization kinetics in proteins  affects the stability of the structure, 
and the aging  of proteins in the cells results in appearance of units of D-amino acids,  and
with the loss of higher order structures \cite{Fujii1}. The homochirality in proteins is in \cite{Tox3} predicted to be
more stable at a smaller water activity, e g. at higher ionic concentrations than in the cytosol solution in biological cells.

$ $\\

TABLE I: Chiral discrimination, $\Delta H_{\textrm{CD}}^{\ominus}(c)$

\begin{tabbing}
\hspace{2.cm}\=\hspace{5.cm}\=\hspace{5.cm}\\
	Molecule \>$\Delta_{\textrm{CD}}H^{^{\ominus}}(c)$ kJ/mol\> References \\
	Alanine   \>   $   $   $   $   $   $   $   $   4.3\> \cite{Contineanu1,Kamaguchi,Silva1}  \\ 
	Valine     \>  $   $   $   $   $   $   $   $  2.4 \>  \cite{Silva2} \\
	Leucine   \>   $   $   $   $   $   $   $   $ 3.4 \>  \cite{Huffman} \\
	Isoleucine     \>   $   $   $   $   $   $   $   $  -5.4 \> \cite{Wu,Vasilev} \\
	Serine \>     $   $   $   $   $   $   $   $   7.4      \>  \cite{Neascsu} \\
	Threonine \>    $   $   $   $   $   $   $   $   -12.0\>  \cite{Contineanu2} \\        
	Proline \>       $   $   $   $   $   $   $   $       3.4\>  \cite{Filipa}  \\        
	Tryptophan   \>    $   $   $   $   $   $   $   $   1.3  \>  \cite{Lukyanova,Kochergina}\\

\end{tabbing}
 
An indirect support of the hypothesis,  that it is the secondary  and tertiary  homochiral structures of a protein  which  ensure the
homochirality, can be obtained from the strength of the chiral discrimination of
the individual amino acids.
An updated table with recent data for the chiral discrimination of crystals of amino acids is given in Table I.
The chiral discriminations,  $\Delta H_{\textrm{CD}}^{\ominus}(c)$,
are obtained from Eq. 1 and  data for   $\Delta H_{\textrm{f}}^{\ominus}(c,L)$ and    $\Delta H_{\textrm{f}}^{\ominus}(c,DL)$. Only
data for less than half of the twenty biological $\alpha$-amino acids exists. But
one can, however, see from the table that most amino acids have
a positive chiral discrimination, i.e. in favor of a racemic crystallization
of the amino
acids. Only two (Isoleucine and Threonine) out of the eight amino acids have a sufficient
negative chiral discrimination to favour homochirality in solid
state. In an aqueous solution the chiral discrimination is, however, less (negative) \cite{Tox3}.

The  homochiral stability of proteins cannot be explained by a  direct  chiral discriminations between units of amino acids
in aqueous solutions, but
 must originate from the $\alpha$-amino acids ability to perform proteins with higher order compact and stable structures, when they are homochiral. 
This property can also explain, why there is no direct relation between the sequence of the units of  L-amino acids
in an enzyme, the secondary structure  and its biological function.   The sequence of the units  in a given enzyme    differs in  biosystems, and
for homologous enzymes these differences can be used to obtain the cladistics of the biological  evolution 
\cite{Margoliash,Roche}. 
The sequence in e.g.  insulin is on one hand specific for a given biological system \cite{Stretton}, but its tertiary conformation,
and thereby the enzymatic function
 in the metabolism is on the other hand maintained to a degree, that   an insulin molecule in one biosystem  can function in  another biosystem.

\section{The establishment and preservation of 
homochiral structures  in carbohydrates.}

Theformation
of a compact hydrophobic core in a protein 
can explain the  stability of homochiral proteins \cite{Tox3}, but one still has to
explain the
formation
and stability of the D- polysaccharides in a prebiotic aqueous environment. 
The central chiral molecule in the formose reaction  for spontaneous synthesis of carbohydrates
\cite{Butlerow} is Glyceraldehyde, and the corresponding key molecule in the Glycolysis is
Glyceraldehyde-3-phosphate. Both molecules have an active isomerization kinetics in an aqueous solution,
\begin{eqnarray}
\textrm{D-Glyceraldehyde-3-phosphate} \rightleftharpoons   
\textrm{Dihydroxyacetone phosphate}  \nonumber\\
\textrm{Dihydroxyacetone phosphate}  \rightleftharpoons  \textrm{L-Glyceraldehyde-3-phosphate}
\end{eqnarray}
and this will, without other chemical constraints, give a racemic composition at
relevant biological concentrations \cite{Tox3}. Furthermore
all living systems contain a very effective enzyme, Triose Phosphate Isomerase,
which eliminates the activation energy in the isomerization kinetics of Glyceraldehyde-3-phosphate
\cite{Alberty}. This
simple enzyme   is believed to have been present in LUCA
(Last Universal Common Ancestor, also called the Last Common Unicellular Ancestor)
and could be among the very first enzymes in the evolution  \cite{Sobolevsky,Farias}.
The enzyme will, without further physicochemical constraints
reinforce the formation of a
racemic composition of carbohydrates in biosystems. So the presence of this enzyme indicates,
that
$\textit{the homochirality of the}$\\
$\textit{carbohydrates in biosystems is not obtained by discrimination, but is esta-}$\\
$\textit{blished and  preserved by another physicochemical mechanism.}$

To understand the formation
and preservation of homochirality  in   polysaccharides and  in the metabolism, one
must notice, that these  chain reactions are reversible.
Take for instant the central step in the Glycolysis,
\begin{eqnarray}
	\textrm{D-Fructose-1,6-diphosphate} \rightleftharpoons  \nonumber\\ 
	\textrm{Dihydroxyacetone phosphate } \nonumber\\
	\textrm{ + D-Glyceraldehyde-3-phosphate}.
\end{eqnarray}   
Although $\Delta G^{^{\ominus '}}$ = 23.97 kJ/mol of this reaction  is strongly positive,
  it readily proceeds in the forward direction under intracellular conditions ($\approx 10-100\mu$M),
and with a Gibbs free energy gain, $\Delta_r G<0$, by cleaving Fructose. But the reverse reaction is spontaneous 
at  higher of  concentrations
of Dihydroxyacetone phosphate  and D-Glyceraldehyde 3-phosphate, and
   with   a gain of free energy by synthesis of Fructose  and a storage of free energy  
in Frucose-6-phosphate, and in the first component, Glycose-6-phosphate, in the Glycolysis.

The isomerization kinetics at the chiral carbon atom in Glyceraldehyde
is, however, not active
after the condensation of Glyceraldehyde and Dihydroxyacetone, and the original chirality  is conserved by the condensation.
The pentoses and hexoses contain additional chiral carbon atoms,
created by this and other synthesis, and with a keto-enol isomerization between them. 
But the chirality at carbon atom No. 5  (No. 4 position for pentoses)
is not affected by these keto-enol isomerizations.
The homochiral structures are not limited to simple D-polysaccharides, also
the chains in RNA and DNA are all D-structures of phosphate esters of Ribose and Deoxyribose, respectively.
D-Ribose-5-phosphate has a metabolism which is linked to the Glycolyse, and altogether these facts can explain the formation and preservation
of a homochiral carbohydrate world:
d.

\begin{figure}
\begin{center}
\includegraphics[width=10cm,angle=0]{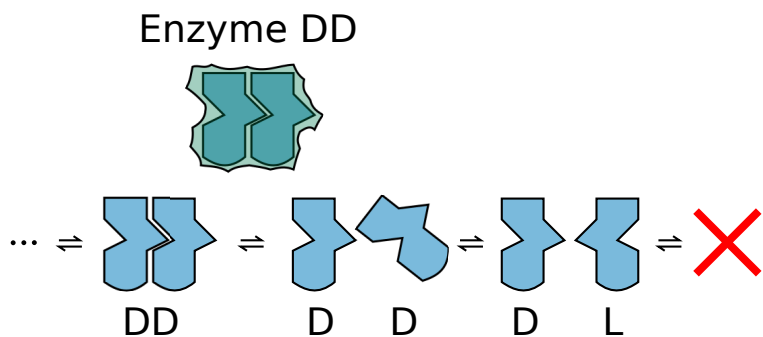}
\end{center}
	\caption{ Illustration of a chain reaction with a stereospecific enzymatic (Enzyme E$_{\textrm{DD}}$) catalyse of one of the steps in
	the chain reaction. The figure illustrates polymerization of a homochiral polymer from chiral monomers with isomerization kinetics and a
	stereospecific catalyse of the first step of the polymerization into  a dimer DD. In the illustration there is
	no enzymes, Enzyme E$_{\textrm{DL}}$ and Enzyme E$_{\textrm{DD}}$, and therefore no polymerization into a non-homochiral polymer.}
\end{figure}

If there at the establishment of the consecutive reactions in the metabolism was an enzyme, which only reduced the activation energy for the
reaction with the D-structure of the carbohydrate, but not for the L-structure,
then a metabolism 
  will  drain the aqueous solution with the  central units, Glyceraldehyde-3-phosphate,  for the  D-conformations.
 The active isomerization kinetic  will convert L-Glyceraldehyde-3-phosphate 
to  D-Glyceraldehyde-3-phosphate   and
  ensure a  racemic equilibrium of the  two enantiomers in the aqueous solution.
  But the net result will be a dominance of D-carbohydrates.
  Such enzymes do in fact exist. The enzyme Hexokinase (Glucokinase),
 which catalyses the first step in the Glycolysis, only catalyses  phosphorylation of  D-Glycose,
 $\textit{but not   L-Glycose}$. The enzyme Ribokinase, which 
 phosphorylates D-Ribose does not catalyse the corresponding phosphorylation of L-Ribose \cite{Chuvikovsky}.
 The first step in the Glycolysis is the stereospecific  phosphorylation of D-Glycose to D-Glycose-6-phosphate. 
RNA and DNA, are  correspondingly phosphor polyesters, so the phosphorylation is central for the establishment of the metabolism as well as the genetics.
An explanation of the role of the phosphorylation in the  Abiogenesis is given by Westheimer \cite{Westheimer}, who argued that the  phosphorylation  stabilizes
the biomolecules. A phosphorylation must have appeared at the establishment of a metabolism. If a stereospecific proteins like
Hexokinase and   Ribokinase by changes were present 
at the prebiotic phosphorylation 
of the carbohydrates,  it is sufficient to ensure the homochirality of carbohydrates in the metabolism.

An illustration of  reactions with chiral objects,  and with isomerization 
kinetics between the chiral units and a  stereoselectiv enzymatic catalyse of
a step in the chain of reactions, is shown in Figure 1. 
The figure illustrates
a polymerization  of chiral units. A homochiral  protein with a  secondary      
structure has a stereospecific surface,
 which can catalyse a chemical reaction. In the illustration there is only a surface which catalyses the D+D  
dimerisation, whereas there is
 no stereospecific surfaces for D+L or L+L  
dimerizations. If a prebiotic environment is enriched with the monomers, 
in the case of
carbohydrates with  Glyceraldehyde by the formose reaction,
and polymerizes, then the isomerization kinetics between the chiral units 
together with a D-stereospecific enzyme somewhere in the chain reactions 
will ensure a dominating D-world.

 A prebiotic  existence of stereospecific proteins  for a reaction in the metabolism 
 can explain the formation and stability of the
 homochiral carbohydrate world. But it also establish an order in the evolution at the start of the Abiogenesis. The creation of a prebiotic  peptide world
 with, by change, a stereospecific  enzymatic  protein for a step in  the metabolism is sufficient for the establishment of  a homochiral carbohydrate world.

\begin{figure}
\begin{center}
\includegraphics[width=10cm,angle=0]{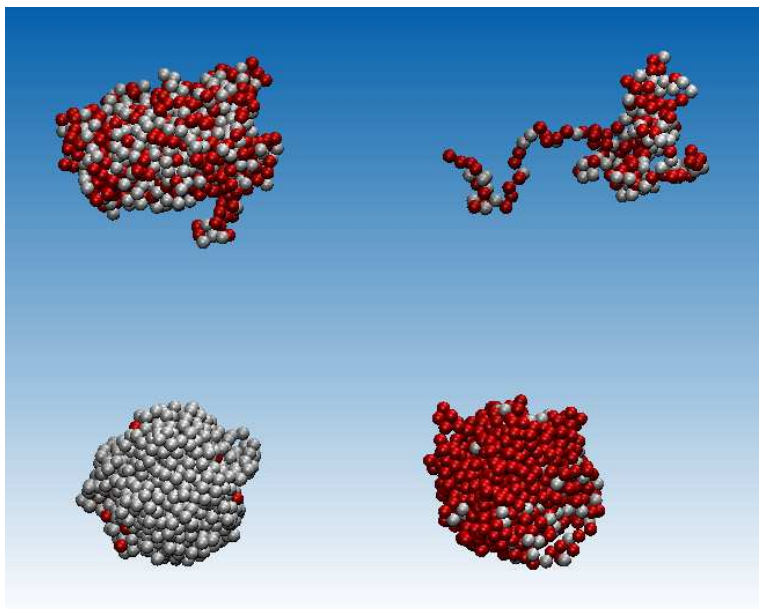}
\end{center}
	\caption{ Illustration of  an aqueous emulsion of proteins. The four  proteins  are from the
	simulations of peptide chains with isomerization kinetics and in aqueous solutions \cite{Tox3}. The activity of water
	in the solution is given by the blue colour (blue: high activity; white: low activity).
	The units of amino acids can have different chirality: red (L) and white (D). The peptide chain
	(upper right)  consists of 200 units and remained
	with a racemic composition  during the simulation. The peptides with longer chains, (1000 units),
	had a racemic distribution of the chiral units for  a high water activity (upper left), but
	for lower activity the isomerization kinetics resulted in compact proteins with  homochiral compositions, but with a statistical equal amount of
 proteins with either D- (lower left)or L- chirality (lower right).}
\end{figure}

\section{The establishment and preservation of 
the L-protein and D-carbohydrate world.}
Linus Pauling's anticlockwise all D-peptide helices  must be the key-molecules in a  D-protein- and L-carbohydrate world,
a world which  not exists on planet Earth.
In \cite{Tox1,Tox2,Tox3} it was pointed out, that one can obtain a spontaneous symmetry break in a racemic system and a dominance of one 
of the  chiral units, when the subdomain of one of the homochiral conformations  percolates the  system. But the compact proteins in an aqueous emulsion are autonomous, and
the isomerization kinetics
can one hand ensure homochirality in the proteins, but will on the other hand give an statistical equal amount of
homochiral proteins with  either D- or L-units. Enzymes are compact proteins with
from 62 to $\approx$ 2500 amino acids. Computer simulations of simple peptide   
chains in an aqueous environment and with isomerization kinetics show,
that short chains remained in  an open structure  with a racemic composition,
whereas longer chains had a compact structure and they were almost homochiral
for low water activity \cite{Tox3}. Figure 2 show four proteins: a relative short
peptide chain with 200 units (upper right)  and  four  proteins with 1000 units.
$\textit{A prebiotic aqueous emulsion of proteins and without a selective me-}$\\
$\textit{chanism will contain both homochiral D- and L-proteins  and not be global}$\\
$\textit{homochiral.}$

There is, however,  a crucial quality of the metabolism  and $\it{catabolism}$ of carbohydrates
and peptides, which   can explain  the emergence and preservation of 
 the L-protein and D-carbohydrate world only. 
The L-carbohydrates are not included in the metabolism of carbohydrates, but simple bacteria  \cite{Shimizu} and procariotes \cite{Yoshida} 
contain  enzymes, which catalyse a catabolic pathway for
L-glycose.
The metabolism of amino acids contains enzymes, which  cleave the peptide bonds, and which are not stereosellective, e.g. Trypsin
and  Pepsin, and enzymes, which include D-amino acids in the catabolism, e.g. D-amino acid dehydrogenase \cite{Olsiewski}.
The critical barrier  for growth and stability of one of the chiral worlds must be the establishment of its metabolism and catabolism.
Who came first will eat the other by the 
catabolism.

\section{The Abiogenesis.}
The Abiogenesis, or the origin of life, is probably  not a result of  a series of single events, but rather the result of a gradual
process with increasing complexity of
molecules and chemical reactions.
This article argues, however,  that there are some "milestones" and order in the evolution of the complex
molecular biostructures and networks of chemical reactions.

One of the elements in the Abiogenesis is the establishment
by time of an aqueous emulsion of homochiral compact proteins. The homochirality can be obtained by the isomerization
kinetics and preserved 
by  the  compact secondary and tertiary structures of the proteins in aqueous emulsions with low 
water activity (e.g. high ionic concentrations) \cite{Tox3}.

Another  link in the evolution is
the establishment of a metabolism. The first step in the Glycolysis is the stereospecific  phosphorylation of Glucose
to D-Glucose-6-phosphate. 
RNA and DNA, are  correspondingly phosphor polyesters, so the phosphorylation is central for the establishment of
the metabolism as well as the genetics.
The synthesis of peptides and carbohydrates  have appeared simultaneously in the prebiotic aqueous environment. 
 A stereospecific phosphorylation can be obtained by proteins, which by change acted as enzymes for phosphorylation
 of D-glucose and D-Ribose only and thereby established the D-carbohydrate world.
Whereas one  today can find a few examples of D-amino acid units in biosystems, L-carbohydrates does not  appear.
The explanation for this lack of L-carbohydrates and dominance of peptides with units of L amino acids today can very well be the same
as the explanation  for the formation and preservation of homochiral carbohydrates given here:
the existence of the stereospecific enzymes like Hexokinase and Ribokinase together with a stereospecific catabolism of
L-carbohydrates is sufficient to establish the homochirality in the metabolism and to remove the L-carbohydrates and proteins with D-units.

Can this order in the evolution  be verified? Perhaps not, the prebiotic world has disappeared on planet Earth.
But there is a chance that frozen ice in cavities in the
rocks beneath the evaporated oceans on planet Mars
contains  prebiotic materials, which can reveal the start of the Abiogenesis.

				\section{Acknowledgment}
Jeppe C Dyre and Ulf R Pedersen are gratefully acknowledged.
This work was supported by the VILLUM Foundation’s Matter project, grant No. 16515

\end{document}